\documentclass[12pt]{article}
\usepackage{pic03}
\usepackage{hyperref}
\usepackage{url}
\usepackage{graphicx,wrapfig}
\usepackage{macros,macros_static}

\begin{document}

\title{
{
\vspace{-7.0cm} \normalsize \hfill
\parbox{48mm}{CERN-TH/2003-229}
}\\[45mm]
\vspace{-0.0cm}
\bf NEW PERSPECTIVES FOR B-PHYSICS FROM THE LATTICE
}
\author{ Rainer Sommer
        \\
{\em DESY, Platanenalle 6, 15738 Zeuthen, Germany }\\
{\em and}
\\
{\em CERN-TH, 1211 Geneva 23, Switzerland }
}
\maketitle

%
%
\begin{figure}[h]
\begin{center}
%
%
%
%
 \vspace{4.5cm}
\end{center}
\end{figure}

\baselineskip=14.5pt
\begin{abstract}
We give an introduction to the problems faced on the way
to a reliable lattice QCD
computation of B-physics matrix elements. 
In particular various approaches for dealing with the
large scale introduced by the heaviness
of the b-quark are mentioned and promising recent achievements
are described. We present perspectives for future 
developments.
\end{abstract}
\newpage

\baselineskip=17pt

\section{B-physics and lattice QCD}
The truly beautiful
results from recent B-physics 
experiments~\cite{pic:bexp1,pic:bexp2,pic:bexp3,pic:bexp4,pic:bexp5}
represented highlights of this conference. 
Some of them, such as the 
$\rm B-\bar{B}$ mass difference $\Delta m_{\rm d}$, require 
knowledge of QCD-matrix elements for their interpretation in
terms of parameters of the standard model of particle physics
and its possible extensions. This motivates investigations
in lattice QCD, our best founded theoretical formulation
of QCD. Indeed, this formulation allows for the
computation of low energy hadronic properties through 
the Monte Carlo evaluation of the Euclidean path integral.
While such a computation necessarily involves approximations,
which we will discuss, the important property of the 
lattice approach is that all {\em approximations} can be {\em systematically 
improved}. 
 
Before going into the details we summarize the goals of lattice QCD
computations with b-quarks. They motivate the considerable effort involved.
\begin{itemize}
  \vspace{-0.3cm}\item	The determination of parameters of the CKM matrix,
	which in the Standard Model are fundamental parameters 
	of Nature.
	In particular the unitarity triangle should be determined
	and over-constrained in order to test the Standard Model.
	The necessary non-perturbative matrix elements should be computed 
	through lattice QCD. 
  \vspace{-0.3cm}\item A precise computation of the b-quark mass, 
	which enters many phenomenological predictions
	and
	plays a r\^ole in grand unification and other questions beyond the 
	Standard Model.
  \vspace{-0.3cm}\item	A determination of the spectrum and lifetimes of b-Hadrons.
  \vspace{-0.3cm}\item	Non-perturbative tests of the Heavy Quark Effective Theory (HQET) 
	which is 
	applied frequently to simplify the dynamics of heavy quarks, 
	but is difficult to test experimentally.
\vspace{-0.3cm}\end{itemize}
 
The starting point of a lattice computation is
the QCD Lagrangian, formulated on the
discretized 4-dimensional Euclidean space-time,
i.e. on a hyper-cubic lattice with spacing 
$a$ \cite{Wilson,books:creutz,books:MM}. The beauty of
this theory is that it contains only the (bare) gauge coupling $g_0$ 
and the (bare) quark masses as parameters. After 
these parameters have been fixed by a small set of experimental
observables, say the masses of proton, pion, kaon, D-meson and B-meson,
all observables 
become predictions of the theory\footnote{For simplicity
we neglect the top quark and assume that electroweak effects 
are treated as perturbations.}. 
In order to have a finite number of variables in the Monte Carlo 
evaluation of the path integral, one considers a finite space time 
of linear size $L$, mostly with periodic boundary conditions.
One then has to approach 
\begin{itemize}
  \vspace{-0.1cm}\item[$\Rightarrow$] the infinite volume limit, $L\to\infty$ and
  \vspace{-0.3cm}\item[$\Rightarrow$] the continuum limit, $a\to0$ .
\vspace{-0.3cm}\end{itemize}
While the statistical errors of the Monte Carlo decrease
$\propto 1/\sqrt{\mbox{computer time}}$ 
at fixed $a$, $L$, a reduction of the 
total error requires to approach the above limits. Therefore the total error 
decreases much more slowly as computers get faster. Progress is made
by better formulations of the problems, the development
of better computational algorithms and the rapid increase in
computer speed.  
Here we want to explain the particular challenges one faces in 
reaching small overall errors in B-Physics
and discuss recent advances in facing them. Some results will
be shown, mainly to illustrate the progress that has been made.
A more comprehensive list of results can be found in 
\cite{lat01:ryan,lat02:yamada}.  
\section{The challenges}
\subsection{Renormalization}
One of the challenges that one faces has to do with the fact
that interesting transitions originate 
from the electroweak interactions, which we can not treat simultaneously
with QCD in the simulations. One then adopts the (good)
approximation to treat the electroweak interactions at the lowest
non-trivial order in the electromagnetic and weak coupling.  
Consequently one has an {\em effective} Hamiltonian,
valid at energies far below the electroweak scale, 
which contains
the quark and gluon fields but not the photon and the electroweak 
bosons. For example a left-left 4-fermion operator
is one of the operators in the effective Hamiltonian
and its matrix element determines $\Delta m_{\rm d}$. 
The renormalization of such operators
is a non-trivial task, but besides perturbation theory \cite{pert:stefano}
powerful non-perturbative approaches have been
developed \cite{RIMOM,alpha:sigma,impr:lett,mbar:pap1} and they
are continuously improved and applied to new operators \cite{lat02:rainer}.
We will come back to this challenge in our discussion of
the HQET.
\subsection{The multi-scale problem}
Even after eliminating the electroweak scale
by considering low energy matrix elements,
b-physics always contains two more scales apart from the 
typical QCD scale $\Lambda= \rmO(1\GeV)$. 
The first is the scale of the light quark masses, or better
the mass of the pion and the second is the large mass of the b-quark
itself. To understand what this means in practical terms,
consider a lattice with $32^3 \times 64$ points, as it is 
presently possible in the quenched approximation but still 
prohibitively expensive for the full theory.\footnote{After
integration over the quark fields the Boltzmann weight of the
path integral contains a factor of the determinant
of the Dirac operator. 
In the quenched approximation the dependence of this determinant
on the gluon fields is neglected. 
In a perturbative language this corresponds to neglecting quark 
loops, but keeping all the gluon exchanges.
}
Choosing $L\approx 2.5\,\fm$ and $a\approx0.1\,\fm$,
one notices that $L$ is comparable to the Compton wave length 
$\mpi^{-1}$, namely $L \approx 2/ \mpi$,
and at the same time the b-quark mass is larger than the 
inverse lattice spacing: $\mbeauty \approx \frac32 a^{-1}$. 
Instead we should have
\bes
  L \gg  \mpi^{-1}\,,  \label{e_L}\\
  a \ll  \mbeauty^{-1} \label{e_a}\,,
\ees
in order to keep finite size effects due to pion
propagation around the periodic world small ( \eq{e_L})
and to properly resolve the propagation
of a b-quark (\eq{e_a}). It is known that effects
of the first type will become rapidly (exponentially) 
small when $L > 4 m_\pi^{-1}$ \cite{finitesize:martin1,GaLe:87},
while the dominant discretization errors due to a heavy
b-quark are given by $\rmO((a \mbeauty)^2)$ if the theory is
$\Oa$-improved (i.e. linear $a$ effects are 
removed \cite{impr:SW,impr:pap1}). The latter behavior
sets in roughly for $a < \frac14 \mbeauty^{-1}$ \cite{zastat:pap2}. 
On a $32^3 \times 64$ lattice one is far from 
satisfying these constraints. An additional very relevant 
factor is that the computational cost of simulating 
full QCD grows rapidly when quark masses are lowered.
This means
that presently simulations at the physical values of the light quark
masses are impossible.

In short: {\em the light quarks are too light and the heavy quarks 
are too heavy for present computing capabilities}. 

In order to obtain results,
a reformulation of the theory (or specific problem) is needed
or extrapolations have to be performed.  

\subsection{Chiral extrapolations \label{s_chir}}
Reliable extrapolations of numerical data
are possible when sufficient analytic knowledge
of the functional dependence on the extrapolation parameter 
exists {\em and when the data are available in a range where the 
analytic formulae apply}. Concerning the light quark mass 
dependence,
chiral perturbation theory furnishes 
an expansion in terms of them,
which is also applicable in the case of heavy-light mesons \cite{chir:fDs1,chir:fDs2}. 
Unfortunately it remains unclear up to now, how small the 
quark mass has to be in order that this expansion is applicable at the 
quantitative level.
Most current simulations reach light quark masses which are about half 
as heavy as the strange quark mass and an agreement with 
the analytic formulae could not yet be established 
\cite{lat02:panel}. 
For recent reviews of the numerical situation 
with an emphasis on B-physics see refs.~\cite{reviews:laurent,pic02:andreas}. 

As an example we would like to mention only that estimates of the
uncertainty for the phenomenologically important ratio
$
  \xi = {(\Fbs \sqrt{\Bbs}) /( \Fbd \sqrt{\Bbd})}
$
range approximately from 5~\% \cite{chiextr:donoghue} to 10~\% 
\cite{chiextr:kronryan}. In our opinion these estimates 
have to be substantiated by simulations with smaller quark
masses, where contact with the chiral perturbation theory
expansion can be demonstrated. 
It appears that this requires both faster computers 
and the development of algorithms which perform better,
in particular at small quark masses. First steps have been taken
\cite{algo:GHMC,algo:GHMC3,lat03:MH}, and new ideas exist 
\cite{algo:schwarz}.
These very important developments are not specific to B-physics. We 
therefore do not discuss them further.

\subsection{Heavy quarks on the lattice.\label{s_hq}}
\begin{wrapfigure}{r}{7.9cm}
  \centerline{\hbox{ \hspace{0.2cm}
    \includegraphics[width=7.5cm]{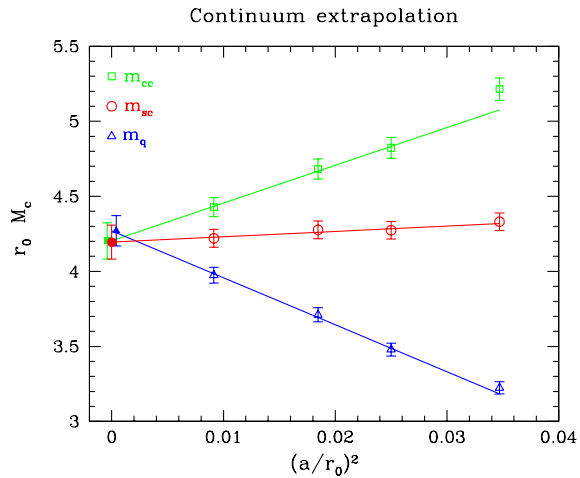}
    }
  }
 \caption{\it
      RGI charm mass from \cite{mbar:charm1}. 
	The length scale $r_0\approx0.5\,\fm$ is derived from
	the force between static quarks.
    \label{f_mcharm} }
\vspace{-0.3cm}
\end{wrapfigure}
Before coming to the methods for dealing with the 
problem of a heavy b-quark, let us give an illustration of the 
extent of the problem. It has been very well investigated 
for the charm quark \cite{mbar:charm1,fds:JR03}, which is a 
factor 4 lighter than the b-quark. In \fig{f_mcharm},
we show the renormalization group invariant (RGI) 
charm mass\footnote{A definition is given in \eq{e_RGI}, below.},
computed on four lattices, between $a=0.1\,\fm$ and $a=0.05\,\fm$. 
Three definitions
of the renormalized quark mass were considered, differing
by discretization errors $\Oasq$. The figure shows 
that on lattices of the chosen size 
(from $16^3\times 32$ to $32^3\times64$)
one can control the discretization errors by extrapolation,
but it is also obvious that this will not be possible 
for much heavier quarks. For b-quarks other approaches have to be 
considered. The following list summarizes what is
being discussed.
\begin{itemize}
  \vspace{-0.3cm}\item[1.] Extrapolation in the mass of the 
	heavy quark, $m_h$. 
  \vspace{-0.3cm}\item[2.] Effective theories which implement an expansion
	in $1/\mbeauty$: NRQCD\cite{nrqcd:first} and 
	HQET\cite{stat:eichhill1}. 
  \vspace{-0.1cm}\item[3.] The ``Fermilab approach'' \cite{fermilabappr}.
  \vspace{-0.3cm}\item[4.] Anisotropic lattices with 
	$a_{\rm time} \ll a_{\rm space}$~\cite{aniso_bphys1,aniso_bphys2}.
  \vspace{-0.3cm}\item[5.] Combinations, in particular of 1. and 2.
  \vspace{-0.3cm}\item[6.] Extrapolation of finite volume effects in the quark 
	mass \cite{mbeauty:romaII,fb:romaII}.
\vspace{-0.3cm}\end{itemize}
1. is clearly applicable
once $m_h$ can be made large enough for the functional form,
a power series in $1/m_h$, to be accurate. 
If $F(m_h,a m_h)$ is the desired observable
at finite lattice spacing and (unphysical) heavy quark mass,
one has to evaluate
\bes
  F(\mbeauty,0) = \lim_{m_h \to \mbeauty} \lim_{a m_h \to 0} 
                        F(m_h,a m_h)
\ees
by two subsequent extrapolations whose order is important. 
In practice a residual
uncertainty remains due to the assumptions made in the
extrapolation $m_h \to \mbeauty$. 
\\
2. Effective theories should be rather accurate since
$ \Lambda / \mbeauty $ is a small 
expansion parameter, which controls both NRQCD and HQET. 
For reasons which will become clear below,
NRQCD has predominantly been used in recent years
although its continuum limit does 
not exist. In this theory one must keep the lattice spacing finite and
control the discretization errors by adding terms to
the Lagrangian that remove them approximately. An additional
source of errors is that the coefficients in the Lagrangian 
are usually determined perturbatively. 
Below we will report on recent progress in HQET on the lattice.
\\
3. The Fermilab approach was discussed at last year's 
conference~\cite{pic02:andreas}.
\\
4. The possibility whether 
anisotropic lattices do permit a situation $a_{\rm space}\mbeauty > 1$
for the interesting matrix elements is still under discussion
\cite{aniso_bphys1,aniso_bphys2,heavyquarks:AKT}. 
In any case, giving up the symmetry between space and time
allows for dimension four
operators in the Lagrangian which break Euclidean invariance.
To obtain a Euclidean invariant continuum limit (and thus Lorenz
invariance after rotation to Minkowski space), these parameters have to
be tuned properly. For full QCD with dynamical fermions, this is a
non-trivial task if non-perturbative precision is desired.
\\
5. Combining the effective theory at the lowest order, 
(the ``static theory''), with (1.) turns extrapolations 
into an interpolations, reducing the influence of assumptions  
concerning the $1/m_h$-dependence considerably.
\\
6. This  
new idea put forward at Tor-Vergata 
\cite{mbeauty:romaII,fb:romaII} will be explained below.

The dominant approach in the last decade has been to compare
several methods with their strengths and weaknesses and
apply the result to phenomenology if different methods agree. 
New developments offer the chance to 
establish precise results without recourse to crosschecks through 
other methods. 
 
\section{New developments}
\subsection{Non-perturbative HQET \label{s_HQET}}

The lattice Lagrangian of (velocity zero) HQET,
\def\vecsig{{\bf \sigma}}
\def\vecD{{\bf D}}                        
\def\vecB{{\bf B}}                        
\bes \label{e_hqetlag}
\lag{\rm HQET} = \heavyb\,D_0\,\heavy
                                - { c_{\rm kin} \over 2  \mbeauty}\heavyb \vecD^2 \heavy
                                - { c_{\sigma} \over 2  \mbeauty} 
                                \heavyb \vecB\cdot\vecsig \heavy+ \ldots 
\ees
has the same form  as the continuum one to order $1/\mbeauty$;
only the definitions of the covariant derivatives $D_\mu$ and
the chromomagnetic field strength $\vecB$ are of course lattice
specific. 
Together with a similar expansion  
of the operators who's matrix elements
one is interested in, it implements a systematic expansion 
in terms of $1/\mbeauty$ for B-mesons at rest \cite{stat:eichhill1}.
Despite its 
attractive feature of having a continuum limit
order by order in the $1/\mbeauty$-expansion, it has not been applied 
very much in recent years. The reason is threefold. 

\begin{wrapfigure}{r}{7.9cm}
  \centerline{\hbox{ \hspace{0.2cm}
    \includegraphics[width=7.5cm]{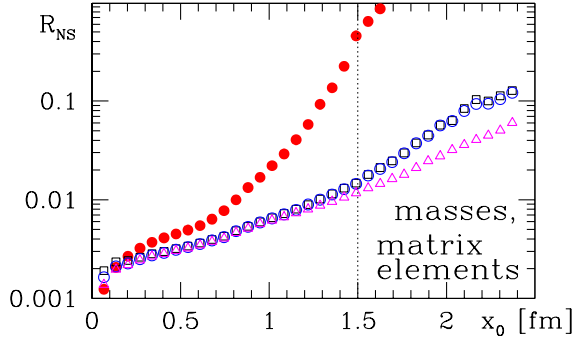}
    }
  }
 \caption{\it
      Noise to signal ratio of a B-meson correlation
	function in static approximation at $a\approx0.08\,\fm$ 
	\cite{stat:letter}. The red bullets are for the original
	Eichten-Hill action while purple triangles correspond to
	the alternative discretization using HYP-links.
    \label{f_rns} }
    \vspace{-0.1cm}
\end{wrapfigure}

First, 
already in lowest order of the effective theory, called the static 
approximation, statistical errors
grow rapidly as the Euclidean time-separation of correlation functions
is made large (\fig{f_rns}, filled symbols). 
But it is in the large time range, 
say $x_0>1.5\,\fm$,
where masses and low energy matrix elements may safely be extracted.

Second, the number of parameters in the effective theory 
grows with the order in the expansion in $1/\mbeauty$.

Third, these parameters have to be determined non-perturbatively;
otherwise the continuum limit does not exist \cite{stat:MaMaSa}
(fine tuning of parameters). This fact is due to the 
mixing of higher dimensional operators, such as
${ 1 \over 2  \mbeauty}\heavyb \vecD^2 \heavy$ 
 with lower dimensional ones, such as $\heavyb\,D_0\,\heavy$. 
A perturbative  
estimate of the parameters in the Lagrangian
(or equivalently of the mixing coefficients) to order  $g_0^{2l}$
would leave a perturbative
\bes
  {\rm remainder} \sim g_0^{2(l+1)}\,a^{-p} \;\sim\; a^{-p}\,
     [\ln(a\Lambda)]^{-(l+1)}
  \;\;\toas{a\to0} \;\;\;  \infty \,.
\ees
This
basic fact is unavoidable in an 
effective theory formulated with a cutoff.  

Recently it was shown that the first point is overcome
by considering alternative discretizations of the static theory
\cite{stat:letter}, which differ only in the way the gauge fields enter 
the latticized covariant derivative. So-called 
HYP-links \cite{HYP,HYP:pot} correspond to the triangles in \fig{f_rns}
and result in errors at the \% level at $x_0=1.5\,\fm$ with an only 
slow growth as $x_0$ is increased. Very importantly, it was also
shown that $a$-effects with this new discretization are small
\cite{stat:letter,lat03:michele}.

The second and third point above can be solved in one go
if  the {\em parameters} of HQET are {\em non-perturbatively} 
determined from those
of QCD. In this way 
the predictive power of QCD is transfered to HQET.

The basic idea how to do this \cite{lat01:rainer,lat02:rainer},
illustrated in \fig{f_mbstrat}, is easily
explained. In a finite volume of linear
extent $L_0=\rmO(0.2\,\fm)$, one may realize lattice
resolutions such that
$\mbeauty a \ll 1$ and
the b-quark can be treated as a standard relativistic fermion.
At the same time
the energy scale $1/L_0=\rmO(1\,\GeV)$ is still significantly below
$\mbeauty$ such that HQET applies quantitatively.
Computing the same suitable observables
in both theories relates the parameters of the HQET Lagrangian
to those of QCD (matching).
One then remains in HQET, changing iteratively to larger and larger volumes,
and
computes HQET observables in each step.
Finally, in a physically large volume (linear extent $\rmO(2\,\fm)$)
the desired matrix elements are accessible. 
At the end any dependence
on the unphysical intermediate volume physics is gone except for terms
of the order $\rmO((L_0\mbeauty)^{-(n+1)})$ if the effective theory was
considered up to order $n$. 

\newcommand{\ftext}[1]{\fbox{ {#1} }}
%

\newcommand{\cred}{}
\newcommand{\cblu}{}
\newcommand{\cmag}{}
\newcommand{\cgre}{}
\newcommand{\cbla}{}

\newcommand{\mgt}{\cmag}

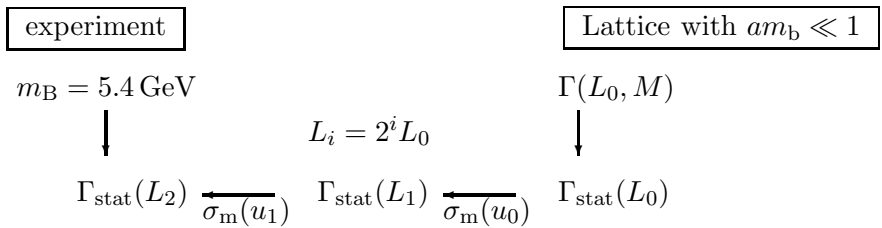
\begin{figure}[b]
\vspace{2.5cm}
\begin{picture}(8,20)(0,0)
\small
  \unitlength 0.4cm
  \put(2,6){\ftext{experiment}}            \put(20.5,6){\ftext{Lattice with 
$a\mbeauty\ll 1$}} 
  \put(2,4){ $\mB=5.4\,\GeV$}    \put(20,4){ $\meff(L_0,M)$} 
  \linethickness{0.3mm}\cgre\put(5.3,3.5){\vector(0,-1){1.5}}
  \linethickness{0.3mm}\cgre\put(21,3.5){\vector(0,-1){1.5}}
  \linethickness{0.3mm}\cbla
  \put(4,0.5){ $\meffstat(L_2)$}
  \put(12,0.5){ $\meffstat(L_1)$}
  \put(20,0.5){ $\meffstat(L_0)$}
  \put(18.9,0.7){\vector(-1,0){2.4}}
  \put(10.9,0.7){\vector(-1,0){2.4}}
  \put(16.5,-0.1){$\sigmam(u_0)$}
  \put( 8.5,-0.1){$\sigmam(u_1)$}
  \put(12,2.5){\small $L_i = 2^i L_0$}
\end{picture}
\caption{\label{f_mbstrat} 
\it
Connecting experimental observables to renormalized HQET.
$\meffstat$ is a renormalized quantity in HQET and $\sigmam(\gbar^2(L))$
connects $\meffstat(L)$ and $\meffstat(2L)$. In the chosen example,
the experimental observable is the mass of the B meson.}
\vspace{-0.1cm}
\end{figure}

The strategy is formulated in such a way that the continuum
limit can be taken in each individual step.
To explain this we take a look at the simple
equation which -- at the lowest non-trivial order in $1/\mbeauty$ 
(static approximation) -- relates the B-meson mass to the  
mass of the b-quark. For definiteness we take 
the RGI quark mass. It is given by the large $\mu$ asymptotics
of the running mass, $\mbar(\mu)$ (in any scheme) via,
\bes
  \label{e_RGI}
  \Mb
    = \lim_{\mu \to \infty}
        \mbar_\beauty(\mu) \left[2 b_0\gbar^2(\mu)\right]^{-d_0/2b_0} \,,
\ees
with $b_0,d_0$ the lowest order coefficients of the beta-function and 
the anomalous dimension of the quark mass, respectively (conventions 
as in \cite{mbar:pap1}). In contrast to
$\mbar_\beauty(\mu)$, the RGI-mass, $\Mb$, is scheme independent.

In static approximation it is related to the mass of the B-meson, $\mB$, via
\def\text#1{\mbox{#1}}
\bes \label{e_master}
  \mB 
      = \underbrace{\Estat - \meffstat(L_2)}_{a\to0 \text{ in HQET}} +
        \underbrace{\meffstat(L_2)-\meffstat(L_0)}_{a\to0 \text{ in HQET}}
       +\underbrace{\meff(L_0,{\cblu \Mb})}_
		{a\to0 \text{ in QCD} } 
       +\rmO(\Lambda^2/\Mb)
\,.
\ees 
Here $\Gamma(L,\Mb)$ denotes the energy of a state with quantum numbers of
a B-meson but defined in a finite volume world of linear extent $L$.
The exact definition of this state \cite{lat01:rainer}
is not important to understand the idea but is quite relevant for
the success of a numerical computation of $\Mb$. 
$\meffstat(L)$ is the same energy but evaluated in static approximation
and $\Estat$ denotes the energy (mass) of a B-meson state in
large volume in static approximation. As mentioned before we have
$L_0 \approx 0.2\fm\,,\; L_2=4L_0$.

To appreciate \eq{e_master}, one should first note that energies in the 
effective theory are related
to energies in QCD by a shift 
$\mhbare$ which is universal in the
sense of being independent of the state. This corresponds
to a term $\mhbare\,\heavyb\,\heavy$ in \eq{e_hqetlag}, which we
have dropped there following standard conventions. Since
the operator $\heavyb\,D_0\,\heavy$ mixes with the lower dimensional one
$\heavyb\,\heavy$ under renormalization, the parameter 
$\mhbare$ is linearly divergent ($\sim 1/a$) and must be determined
non-perturbatively. Its universality means
\bes
  \mB &=& \Estat + \mhbare\,, \label{e_mB} \\ 
  \meff(L,{\cblu \Mb}) &=& \meffstat(L) + \mhbare \,, \label{e_meff}
\ees
with one and the same $\mhbare$ at fixed $a$. 
One may now use \eq{e_meff} to determine the parameter $\mhbare$
in the effective Lagrangian from QCD and then insert it
into \eq{e_mB} to determine $\mB$. This represents the 
general logics for obtaining results in the HQET. 
In order to arrive at the continuum limit of the
prediction, one groups terms as in \eq{e_master}, where 
$\mhbare$ drops out of the energy differences 
$\Estat - \meffstat(L_2)$ and $\meffstat(L_2)-\meffstat(L_0)$
and the continuum limit can be taken separately
for each of the terms as indicated. 
\begin{figure}[tbp]
  \centerline{\hbox{ \hspace{0.0cm}
    \includegraphics[width=11.5cm]{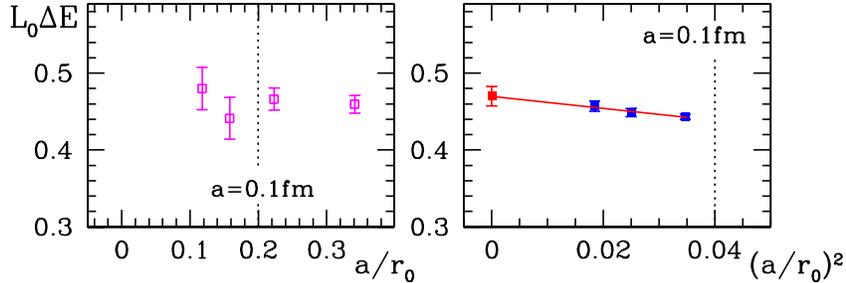}
    }
  }
 \caption{\it
        The combination $\Delta E=\Estat - \meffstat(L_2)$ as
	a function of the lattice spacing. 
	Results with the new discretizations \cite{stat:letter} 
	(r.h.s) are compared with old results \cite{np_hqet:pap1}.
    \label{f_DeltaE} }
\end{figure}
Indeed, in the quenched approximation a precise
continuum limit
has been taken for all terms \cite{lat01:rainer} except for 
the energy difference $\Estat - \meffstat(L_2)$. Here
\cite{np_hqet:pap1} uses values for $\Estat$ from the literature
(l.h.s. \fig{f_DeltaE}). With the new discretization \cite{stat:letter}
more precise results are obtained (r.h.s. \fig{f_DeltaE}),
and they also have the linear $\Oa$-errors removed. 
The full analysis with the new data has not yet been performed.
Using for the time being the value of
$L_0(\Estat - \meffstat(L_2))=0.46(5)$  one 
obtains \cite{np_hqet:pap1} 
(we do not distinguish $\rmO(\Lambda/(L_0\Mb))$ and $\rmO(\Lambda^2/\Mb)$ )
\bes \label{e_mbres}
  r_0\Mb=16.12(28) + \rmO(\Lambda^2/\Mb) \; \to \;
  \mbar_\beauty^\msbar(\mbar_\beauty^\msbar)=4.12(8) \GeV + \rmO(\Lambda^2/\Mb)
\ees
in the quenched approximation and with $r_0=0.5\,\fm$. It is worth 
emphasizing that such a result is based on the non-trivial relation 
between the bare quark masses on the lattice and the RGI-masses
established in \cite{mbar:pap1,impr:babp}. 
\subsection{Results for $\Fbs$ \label{s_fbs}}
\begin{figure}[htbp]
  \centerline{\hbox{ \hspace{0.2cm}
    \includegraphics[width=8.5cm]{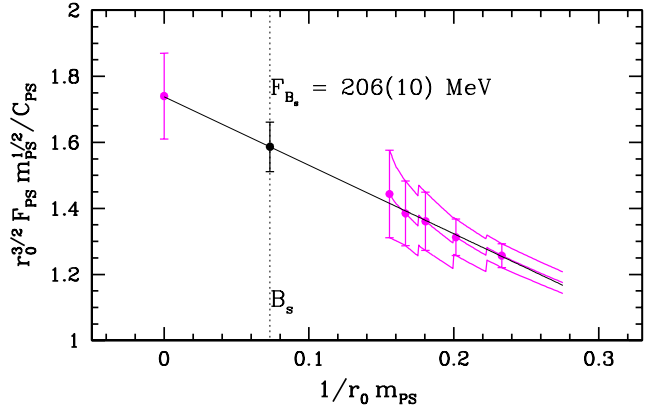}
    }
  }
 \caption{\it
        The continuum limit 
	quenched pseudo scalar heavy-strange decay constant 
	as a function of the inverse heavy-strange meson mass, $\mp$
	\cite{lat03:juri}.
    \label{f_fbs} }
\end{figure}

As a further new development, we show in \fig{f_fbs} a recent computation
of the decay constant of the $\rm B_{\rm s}$ meson in quenched 
approximation \cite{lat03:juri}
using method number 5 in our list of \sect{s_hq}. HQET predicts the mass
dependence
\bes
  \Fp \sqrt{\mp} / \Cps(M/\Lambda) = 
 \PhiRGI + \rmO(1/\mp) \,,
\ees
of the decay constant $\Fp$,
where $\PhiRGI$ is independent of the heavy quark mass.
It has recently been
computed in the continuum limit of the static approximation 
\cite{stat:letter}. 
The factor $\Cps$ is a function of the ratio of the RGI mass of the 
heavy quark and the QCD Lambda parameter. It is now known quite 
accurately from
perturbation theory due to the 3-loop result of
\cite{ChetGrozin}. Also the numbers
of $\Fp$ at finite quark mass shown in \fig{f_fbs} 
have been extrapolated to the continuum limit \cite{lat03:juri,fds:JR03}.
While the subsequent interpolation
in $1/\mp$ is obviously safe, an extrapolation
with just results for $1/(r_0\mp)>0.15$ would depend on
the functional form assumed. The present result
$\Fbs=206(10)\,\MeV$ \cite{lat03:juri} is in good agreement with the recent
one from the Tor-Vergata group \cite{fb:romaII}. 
\subsection{The Tor-Vergata approach: \\
	Extrapolation of finite volume effects in the quark mass
	\label{s_romaII}}
The starting point of this method is the same as in
\sect{s_HQET}: in an intermediate volume (e.g. of size $L_1^3 \times 2L_1$)
the lattice spacing can be made 
small enough to be able to treat b-quarks as relativistic fermions. 
The decay constant $F(M_h,L_1)$ can then be computed for $M_h=\Mbeauty$.
While in the non-perturbative HQET, this was used to 
obtain the parameters in the effective theory, de Divitiis et al. 
remain in the relativistic theory, but introduce as their central observables
finite size effects $\sigma_{\rm F}$ in the following way\footnote{ For $4 L_1 \approx 1.6 \,\fm$ 
further finite size effects are expected to be very small
(at least in the quenched approximation).} 
\bes \label{e_romaII}
 F({ \Mbeauty},\infty) \approx F({ \Mbeauty},4 L_1)  
                               =\underbrace{F(\Mbeauty,4 L_1) \over F(\Mbeauty,2L_1)}_{=\sigma_{\rm F}(\Mbeauty,2L_1)}      
                        \times\underbrace{F(\Mbeauty,2 L_1) \over F(\Mbeauty,L_1)}_{
=\sigma_{\rm F}(\Mbeauty,L_1)}
			\times F({ \Mbeauty},L_1) \,.
\ees
The idea is that the finite size effects $\sigma_{\rm F}$
depend strongly on the dynamics of the light quark,
but if $M_h$  is large enough, they hardly depend 
on that variable. It is thus expected that
finite volume effects can smoothly be extrapolated in the 
heavy quark mass, 
$\sigma_{\rm F}(\Mbeauty,L_1)=\lim_{M_h\to\Mbeauty}\sigma_{\rm F}(M_h,L_1)$. 

\begin{figure}[htbp]
  \centerline{\hbox{ \hspace{0.2cm}
    \includegraphics[height=4.5cm]{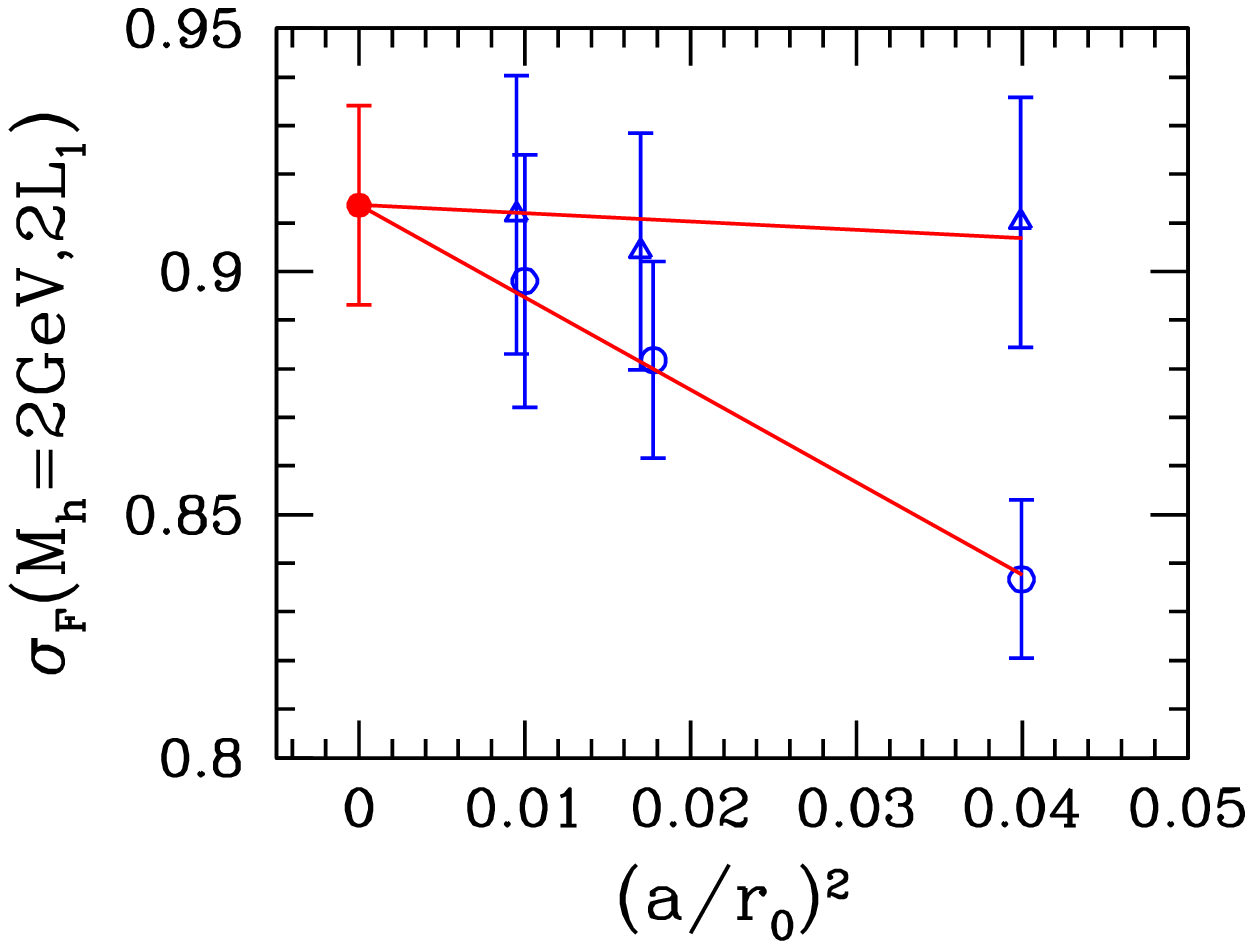}
    \hspace{0.5cm}
    \includegraphics[height=4.5cm]{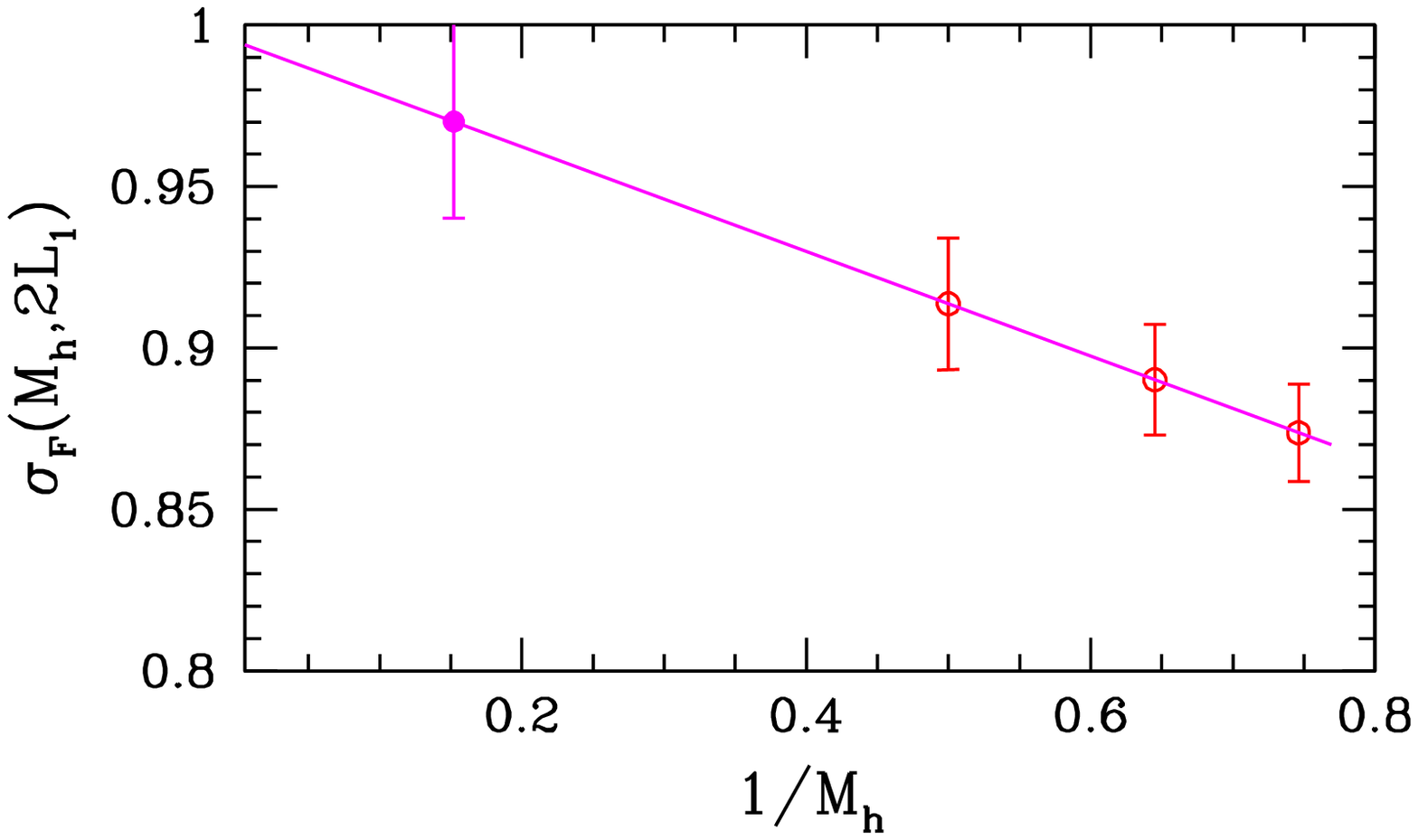}
    }
  }
 \caption{\it
      Quenched computation of $\sigma_{\rm F}(\Mbeauty,2L_1)$ with the
      light quark mass equal to the strange quark mass. 	
    \label{f_romaII} }
\end{figure}
Setting $L_1\approx 0.4\,\fm$, a numerical computation was performed 
in the quenched approximation.
Of course, the three factors in \eq{e_romaII}
are obtained by an extrapolation to the continuum limit at fixed
$M_h$, followed by an extrapolation in $M_h$ to $\Mbeauty$ for 
$\sigma_{\rm F}$, while $F({\Mbeauty},L_1)$ could be determined 
directly at $M_h=\Mbeauty$. 
The most difficult extrapolations are those concerning the 
step $2L_1 \to 4L_1$, since there $a \geq 4L_1/32 \approx 0.05\,\fm$
on lattices with maximally $32^3\times 64$ points. They are shown
in \fig{f_romaII}.
 
Comparing the r.h.s. of \fig{f_romaII} with \fig{f_fbs}
one notices that (looking at the central values) in both cases the
last point of the finite mass numbers differs from
the result at $\mp=\mbs$ by about 7\%. 

The final result is \cite{fb:romaII}
\bes
  \Fbs=192(6)(4)\,\MeV\,.
\ees
Applying the same
method also to the computation of the 
b-quark mass, yields \cite{mbeauty:romaII} 
$
  \mbar_\beauty^\msbar(\mbar_\beauty^\msbar)=4.33(10)\,\GeV
$.

\subsection{Summary}
The new developments discussed in this section 
attack the problem of a heavy quark mass in ways where
all necessary renormalizations are performed non-perturbatively 
(only in $\Cps$ an $\rmO(\alpha(\mbeauty)^3)$ uncertainty remains)
and the {\em continuum limit} can be taken in each step. They have
been shown to be applicable in numerical computations,
yielding good accuracy after propagating errors through all the steps. 
In fact, all the errors quoted in this section can be further reduced in the 
quenched approximation.

\section{Perspectives}
We start with some possible straight forward  
improvements of the above results. \\
First, the precision in \sect{s_HQET} and \ref{s_fbs}
can easily be improved. Once this is done, one may quantify the size 
of $1/\mbeauty$
corrections rather precisely. In this context we note
that it is surprising that 
so far there is no sign of $(1/\mcharm)^2$ terms in the charm region
(see \fig{f_fbs} and \fig{f_romaII}). Better precision 
is desirable to pin those down. \\
Second, a very promising improvement would be to combine
the method of \sect{s_romaII} with \sect{s_HQET}: the finite size effects 
$\sigma_{\rm F}$ can be computed in the HQET. Then the 
mass-extrapolation in \fig{f_romaII} will also be turned into
an interpolation and even larger precision and confidence can be achieved.

In our opinion it is most important (and more ambitious) 
to extend the calculations
in the HQET to the level of including $1/\mbeauty$ corrections.
We do not see any major obstacles on the way to reaching this goal.
The severe problem of power divergences can be overcome by the
non-perturbative matching of QCD and HQET sketched in \sect{s_HQET}. 
In contrast to the other methods discussed, the consequent use
of an effective theory to describe the b-quark is a way
to entirely eliminate the mass scale of the b-quark
and thus the necessity of using large lattices. This then opens
the possibility to obtain results in full QCD,  once the usual
(and large) problems with dynamical fermions, which have nothing
to do with the heaviness of the b,  are solved. 
Indeed, as emphasized in \sect{s_chir} we finally do need results
with light dynamical quarks in order to apply them 
to phenomenology with confidence. 
\\
{\bf Acknowledgments.}
I am indebted to my collaborators in the ALPHA-collaboration
for the possibility to present new results at this conference,
prior to publication. 
I thank the Tor Vergata group for communications
on their approach and for sending the data in \fig{f_romaII}. Furthermore 
I am grateful to the CERN theory division for 
its hospitality and L. Giusti for comments on the manuscript.


\end{document}